# The wealth of nations and the health of populations
## A quasi-experimental design of the impact of sovereign debt crises on child mortality

by

Adel Daoud[1,2,3]*

1. Department of Sociology and Work Science, University of Gothenburg, Gothenburg, Sweden
2. Center for Population and Development Studies, Harvard University Cambridge, The United States
3. The Data Science and AI division, Department of Computer Science and Engineering, Gothenburg, Sweden

* Correspondence: adaoud@hsph.harvard.edu

**Abstract**
The wealth of nations and the health of populations are intimately strongly associated, yet the extent to which economic prosperity (GDP per capita) causes improved health remains disputed. The purpose of this article is to analyze the impact of sovereign debt crises (SDC) on child mortality, using a sample of 57 low- and middle-income countries surveyed by the Demographic and Health Survey between the years 1990 and 2015. These surveys supply 229 household data and containing about 3 million childbirth history records. This focus on SDC instead of GDP provides a quasi-experimental moment in which the influence of unobserved confounding is less than a moment analyzing the normal fluctuations of GDP. This study measures child mortality at six thresholds: neonatal, under-one (infant), under-two, under-three, under-four, and under-five mortality. Using a machine-learning (ML) model for causal inference, this study finds that while an SDC causes an adverse yet statistically insignificant effect on neonatal mortality, all other child mortality group samples are adversely affected between a probability of 0.12 to 0.14 (all statistically significant at the 95-percent threshold). Through this ML, this study also finds that the most important treatment heterogeneity moderator—in the entire adjustment set—is whether a child is born in a low-income country.

# Introduction

The wealth of nations and the health of populations are intimately associated (Banerjee and Duflo, 2012; Deaton, 2015; Preston, 1975; Sen, 2006). Countries with higher economic output tend to have populations with better health outcomes. Healthy individuals tend to have a stronger capability to produce goods and services of higher economic value than less healthy individuals (Burgard and Kalousova, 2015; Catalano et al., 2011; Nandy et al., 2016; Reddy and Daoud, 2020). However, although the association between health and economic prosperity is strong, the extent to which economic prosperity causes improved health remains a disputed issue (Baird et al., 2010; Biggs et al., 2010; Catalano et al., 2011; Conklin et al., 2018; Daoud, 2017, 2015; Daoud et al., 2019b; Daoud and Nandy, 2019; Ferreira and Schady, 2009; Halleröd et al., 2013; Ponce et al., 2017; Preston, 2007). This issue springs from two sources, fueling this dispute.

First, the possibility of reversed causality makes it challenging to determine the order and magnitude of the causal effect. That is, to what extent does health cause better prosperity versus prosperity, causing better health. Scholars have proposed a variety of approaches to handle this issue, using a variety of quasi-experimental designs (Burgard and Kalousova, 2015; Catalano et al., 2011; Ferreira and Schady, 2009). For example, Baird et al. (2010) used a design where the focus is to estimate short-term fluctuations in per capita GDP on infant mortality. Using household data, their study handled reversed causality and thus estimating an adverse causal effect of loss of economic output on infant mortality. A critical causal assumption of this design is that while aggregated infant mortality correlates with aggregate economic prosperity, each individual child's risk of mortality has a weaker correlation. Thus, this weak correlation is taken as a signal of independence between the treatment assignment and outcome. Nonetheless, as the majority of studies use GDP per capita as a proxy for economic prosperity, they often analyze short-term variation in GDP (Harttgen, 2018; Preston, 2007; Vollmer et al., 2014, p. 20). As this variation has an unclear source, a more robust research design would be to use a well-defined exposure of increase or decrease in economic prosperity (Daoud et al., 2017, 2019b; Daoud and Johansson, 2020; Daoud and Reinsberg, 2018).

Second, because most research designs rely on linear models, it remains unclear to what extent the effect found is biased due to model uncertainty—that is, a scholar's choice of parametric specification (Gelman, 2006; Imai et al., 2011; Young, 2009). Even if a quasi-experimental design may help in identifying the order of causality (Imbens and Rubin, 2015), the relationship between the treatment, $D$, and the outcome, $Y$, may not follow the chosen parametric specification (Athey et al., 2019; Daoud and Johansson, 2020; Pearl, 2009). Due to the convenience of linear models—most notably ease of statistical interpretation—most scholars rely on linear models. Yet if $D$ and $Y$ are not linearly related, results will be biased. Despite that scholars use polynomials, logarithmic, and other transformation to capture possible nonlinearities, these parametric specifications may instead overfit the data, hampering external validity (Daoud et al., 2019a; Künzel et al., 2018).

Third, as most studies focus on estimating average effects, to what extent increasing or decreasing economic prosperity creates treatment heterogeneity in health, conditional on social and political context, remains mostly unexplored (Goldthorpe, 2015; Grimmer et al., 2017). Although many population studies analyze treatment heterogeneity using a parametric statistical approach, there remains a vast unexplored space of all possible combinations of heterogeneity (Daoud and Dubhashi, 2020).

The purpose of this article is to address these three gaps jointly by analyzing the impact of sovereign debt crises on child mortality. This purpose distills into two aims.

The first aim is to estimate the average treatment effect of sovereign debt crises (SDC) on child mortality. We measure child mortality using six health outcomes: neonatal (within a month of birth), under-one (infant), under-two, under-three, under-four, and under-five years. An SDC is an event of sovereign default, a government failure, to private creditors, or debt restructuring (Laeven and Valencia, 2013). Instead of analyzing fluctuations in economic prosperity measured by GDP per capita—the most common approach—this study relies on SDC as a quasi-experimental event. An SDC causes a sudden shortfall in economic prosperity that is, by definition, beyond the control of any individual household. As GDP per capita relies on each individual's productivity, this measure depends on the productivity of a woman giving birth to a child. A design relying on such an exposure (GDP per capita), is more likely to create dependence on the outcome (a women's risk of losing a child during birth or within a child's first five years of living), as it is more vulnerable to unobserved confounding than a design that uses an independent exposure on the outcome (Dunning, 2012). Consequently, an SDC design is less vulnerable to unobserved confounding than a design relying on GDP per capita.

While GDP per capita measures the continuous fluctuations of both negative and positive economic prosperity, our SDC design is limited to a sudden shortfall in economic prosperity. An SDC causes a median GDP output loss of 40 percent (Laeven and Valencia, 2013, p. 252), and thus, such a large loss of economic prosperity is expected to cause a sizeable adverse health effect on children's health outcomes. Although our SDC design provides a better causal identification than a GDP design, our design is limited to a government's failure to manage public finances and not the entire span of positive and negative variations of economic prosperity. Nonetheless, as many governments—in both affluent and less affluent countries—have large public debts and are therefore heavily engaged in reducing these debts, our SDC analyses quantify what the impact on children's health outcomes may be when government officials fail to uphold prudent macroeconomic policies (Primo Braga and Vincelette, 2010).

The second aim is to quantify the full response space of SDC-induced treatment heterogeneity in child mortality, conditional on a set of variables. By "full response space," I mean quantifying the extent of *all* treatment heterogeneity, and not only a set of selected interaction models. Defined in the Method section, this set ranges from household to country-level characteristics. This study uses novel methods from computer science that detects treatment heterogeneity inductively (Athey et al., 2019).

## Method

### Data

Based on a sample of 57 low- and middle-income countries surveyed between the years 1990 and 2015, this study uses 229 household surveys from the Demographic and Health Survey (DHS). These surveys focus primarily on measuring the health and living conditions of women and their children. Together with DHS surveyors, women reply to a questionnaire asking them about their health, including a set of questions about their children alive and deceased. From these replies, we create birth and mortality histories of children (Baird et al., 2010). A birth history identifies a child's birth date, whether the child was born alone, and the number of preceding and succeeding siblings.

From these histories, we create subsamples for analyzing the impact of SDC on child mortality at six thresholds: from conception (assumed to 9 months before birth) to 28 days, 12 months (u1), 24 months (u2), 36 months (u3), 48 months (u4), and 60 months (u5). The corresponding child samples are: $n_{neo} = 494\,780$, $n_{u1} = 1\,131\,850$, $n_{u2} = 1\,657\,261$, $n_{u3} = 2\,071\,625$, $n_{u4} = 2\,461\,628$, and $n_{u5} = 2\,833\,711$. These samples contain deceased and alive children, as reported at the time of the DHS interview. Thus, the outcome variable $Y$ is a binary indicator variable of 1 if a child died within the reported threshold and 0 otherwise. From the DHS, we collect a set of controls, $X$, about women's living conditions, including their age, education level, and residential area.

A child, $i$, is exposed to an SDC event, $D_{it}$, if this event occurred within the period of conception to its age threshold. We denote this period with the index $t$, and thus, this treatment $D_{it}$ is an indicator variable taking the value "1" if an SDC occurred and "0" otherwise. We collect SDC data from Laeven and Valencia (2013). They build their data based on information from various sources, including IMF Staff reports.| The year of SDC is the reported year of default to private creditors or the year of debt restructuring. Their data identifies 79 SDC events from 1970 to 2011. Complementing this data, we add a set of country-level controls such as GDP and population size to our adjustment set $X$.

**Statistical Model**

Based on the two aims of our study, our study targets two estimands. The first is the average treatment effect (ATE), corresponding to the following quantity, $E[Y(1) - Y(0)]$. Using the potential outcomes framework (Imbens and Rubin, 2015), the variable $Y(1)$ is the potential-mortality outcome of children under SDC exposure, and $Y(0)$ is the potential-mortality outcome of the same children in the absence of SDC. Thus, the target is to calculate the difference between these potential outcomes in expectations. We identify the ATE mainly on a quasi-experimental design, reinforced by an adjustment set (conditioning on observables). Following the quasi-experimental design proposed by Baird et al (2010), an economic shock is plausibly as-if randomly assigned to each child during the period $t$ (in utero up to the previously specified age thresholds). In other words, no pronounced unobserved confounding exists that both affect an SDC and a child's probability of mortality, beyond the defined adjustment set $X$. For example, as the characteristics of low-income countries do affect the likelihood of SDC and a child's probability of mortality, we include those characteristics in the adjustment set $X$ (mainly GDP and population size).

When relying on our quasi-experimental moment and conditioning on this adjustment set, we gain conditional ignorability by assumption, that is, $Y(1), Y(0) \perp\!\!\!\perp D | X$. Thus we identify the ATE in the following equation,

$$\tau = E[Y(1) - Y(0) \mid X] = E[Y \mid D = 1, X] - E[Y \mid D = 0, X]$$

The first equality defines our estimand, $\tau$. Because we condition on $X$, we may bring $D$ into the equation. Assuming consistency—that $Y = Y(1)$ for SDC exposed children $D = 1$, and $Y = Y(0)$ for unexposed children $D = 0$—we can substitute potential outcomes with factual ones (Imbens and Rubin, 2015). As the sample of SDC exposed children who also died is small, our model weights $\tau$ by the propensity score of SDC exposure (Crump et al., 2009). This score is defined in the quantity, $e(X) = E[D \mid X]$. See Table 1 showing this sample by country that had an SDC overlapping with a mother carrying a child in her womb, and that child subsequently died.

[Table 1 about here]

Our second estimand is the conditional average treatment effect (CATE) that is defined in the following quantity,

$$\tau(x_i) = E[Y(1) - Y(0)|X = x_i]$$

This quantity is identified using the exact same assumptions of our quasi-experimental design, conditional ignorability, and consistency.

To estimate $\tau$ and $\tau(x_i)$, I use the generalized random forest (GRF). The GRF adapts the family of random forest (RF) estimators (Breiman, 2001) for efficient non-parametric estimation of causal effects (Athey et al., 2019). RF models learn ensembles of regression (or classification) trees, each fit a different resampled population and covariate set, to estimate and reduce model variance. Each tree learns a set of rules (e.g., $mother_{age} > 30$) which partition the population of units (children) into different leaves of the tree. The predicted outcome for a new unit is the average of outcomes for observed units assigned to the same leaf; the prediction of the forest is the average of the predictions of all trees. A strength of non-parametric (or machine learning) estimators such as RF is that they are designed to optimize predictive accuracy on held-out data by trading off bias and variance through regularization, rather than learning the parameters of a fixed-size model (Hastie et al., 2009). For tree-based estimators, many heuristic regularization strategies exist, including limiting the depth or number of leaf nodes in each tree. Generally, growing more trees is preferable for out-of-sample generalization. In order to select such tuning parameters, I use sample splitting, evaluating the predictive accuracy on a randomly subsampled set of held-out data, never exposed to the model. GRF uses a version of sample splitting called "out-of-bag predictions." As it randomly picks a subset of cases from the full sample—hence, the name *random* forest—to grow each tree, it does not use all cases for all trees. Out-of-bag prediction applies each specific case to only those trees in which the GRF did not sample that case to grown those trees. This type of prediction is a more efficient way to use data.

Once the GRF produced $\tau$ and $\tau(x_i)$ estimates, the population may be clustered at different levels of granularity to compile group-specific average treatment effects. GRF uses clustered-robust errors that ensures that the standard errors are computed correctly and that they are less sensitive to outliers. In our study, the SDC exposure occurs at the country-year level. For example, because two alive children of the same country and DHS sampling year (e.g., the year 2000) may be born in different years, say one child in 1995 and the other in 1996, these two children have a different SDC exposure probability. For under-five mortality, an SDC occurring anytime between 1994 (counting the period in utero) to 2000 is a valid exposure, yet the older child has a higher propensity of exposure than the younger child. However, as the SDC is not an individual-level event, we also want to correct for country-year clustering. A country-year clustered standard errors adjust for variation in exposure and for this clustering.

## Results
While Table 1 shows the number of children that died concurrently with an SDC, Figure 1 shows the entire distribution of children born in SDC plagued countries. The blue bars show the frequency of children that remained alive, and the red bars children that died. The x-axis is normalized by when an SDC occurred (most countries in our sample had only one SDC), signified by the red line centered at x equals 0. Children's birth year is then normalized by counting from an SDC event. The dashed bars show when a DHS survey occurred. Taking all this information, this figure shows that our effective sample size of children exposed to an SDC is reduced to only those children born near the red line. Despite that this quasi-experimental design reduced our sample drastically, our analysis focuses only on the part of the variation that is well identified to estimate $\tau$ and $\tau(x_i)$.

[Figure 1 about here]

The GRF estimates a large adverse SDC average effect on child mortality, $\hat{\tau}$, as shown in Figure 2. While SDC adversely increases the probability of mortality for a newborn child (neonatal mortality) by 0.02, this effect is not statistically significant. The sample size of women experiencing an SDC and conceiving a child, sometime between inception to the first month of newborn's life is small (n = 14 605). Compare this with the sample size of the same group of women but those that did not experience a SDC (n = 480 175). Consequently, our neonatal GRF estimation is likely underpowered—it lacks a sufficiently large sample size—to answer whether SDC impacts neonatal mortality.

[Figure 2 about here]

An SDC adversely affects child mortality at the other age thresholds as well, and this effect is statistically significant beyond the 95-percent level. These adverse effects are also comparable in magnitude. For example, while infant mortality increases by a probability of 0.12 (std 0.059), under-two mortality increases by 0.14 (std 0.06). The probabilities of under-three, -four, and -five mortality lies between the probability interval of 0.12 and 0.14.

Around $\hat{\tau}$, our GRF estimates a considerable variation in the CATE, $\hat{\tau}(x_i)$. As shown in Figure 3, treatment heterogeneity has a different distribution between the different mortality cohorts. The blue line show $\hat{\tau}$. At zero effect (red line), there is a large hump of children that have zero SDC effect on their probability of mortality. Analyzing the characteristics of these children, I find that many of these children are living in countries that normally do not have an SDC.

[Figure 3 about here]

For neonatal mortality (panel A)—and to some extent, under-two mortality—treatment heterogeneity is bimodal. One mode positions itself around the red-line (zero effect), and the other model centers around an SDC effect of 0.09. To identify which characteristics moderate this bimodality, the GRF produces a variable importance plot. Variable importance is a measure of how often the GRF use a variable to grow a tree during the estimation procedure. The more often it uses a variable, the more treatment heterogeneity it likely carries. For example, a measure of 0.2 indicates the proportion of times the GRF used a variable to grow a tree. Figure 4 shows this plot, and the three most important variables are the size of the country population ("pop"), economic development ("rgdpna"), and mother's age ("mother_age") when giving birth to the child. Characteristics that normally appear in other interaction studies, such as the child's gender, age, residential area, and mother's education level, are less important moderators.

[Figure 4 about here]

## Discussion and Conclusion

While economic prosperity and health outcomes are strongly associated, the direction of the causal effect remains challenging to identify (Banerjee and Duflo, 2012; Deaton, 2015; Preston, 1975; Sen, 2006). To what extent does variations in prosperity causal changes in health outcomes, and vice versa? As most studies rely on GDP per capita to measure economic prosperity, disentangle this direction will likely remain a challenge. To identify a part of this causal variation, this study relies instead on the quasi-experimental nature of economic downfall caused by a sovereign debt crisis (SDC), and its effect on child mortality. Because an SDC event is independent of a child's risk of mortality—especially in the first couple of months of a child's birth—SDC provides a quasi-experimental opportunity to quantify the effect of loss of economic prosperity on a key health outcome.

For children older than one year, this study finds that an SDC causes an adverse average increase of about 0.13 probability. For example, in a population of 100 00 children and in the event of a sovereign default, our statistical estimation shows an increase of excess mortality of 13 000 children. This estimation is considerable. Nonetheless, as an SDC causes a median GDP output loss of 40 percent (Laeven and Valencia, 2013), such a large loss of economic prosperity is expected to cause a large adverse effect on children's health outcomes.

Although the SDC effect on neonatal mortality is zero, we find considerable heterogeneity driven by middle- versus low-income countries separating the effect between middle-income countries at zero and low-income around an effect of 0.09 probability increase in mortality. As most studies using parametric models focus on testing moderation in a child's gender, age, residential area, and mother's education-level (Burgard and Kalousova, 2015; Catalano et al., 2011), this moderating result has been largely overlooked as being the most important one of all in the adjustment set.

Our findings contribute to our knowledge on the relationship between economic decline and child health (Banerjee and Duflo, 2012; Burgard and Kalousova, 2015; Daoud, 2018, 2011, 2007; Deaton, 2015; Preston, 1975; Stuckler and Basu, 2013). In their review of the economic crisis of 2008, finds that "many other suspected associations remain poorly studied or unsupported. The intuition that [infant] mortality increases when the economy declines, for example, appears wrong." (Catalano et al., 2011, p. 431). Using a large sample of birth history, this study finds strong empirical support that economic decline increases the probability of child mortality, for children between the ages of one and five. However, a noticeable difference between most studies and our study is using SDC instead of GDP.

Our study has one notable limitation. Despite the quasi-experimental design and the adjustment set, unobserved confounding may still bias the results. Our statistical model adjusts for a variety of observed covariates, yet data on the behavior of governments is lacking. For example, I adjust for the fact that countries with weak governance and unstable institutions are both more likely to default, and children more likely to die at a younger age. However, some governments in low-income countries have been found to seek an SDC to access desirable loans from international financial institutions such as the World Bank and the International Monetary Fund (Daoud et al., 2020; Stuckler and Basu, 2013). Another possible unobserved confounder is exogenous natural or political shocks that erode public finances, pushing a government into a default (Coutts et al., 2019; Daoud et al., 2016). Future studies may handle the challenges posed by such confounding.

# Table and Figures

## Tables

| | Country | freq |
|---|---|---|
| 1 | Angola | 37 |
| 2 | Bolivia | 460 |
| 3 | Brazil | 189 |
| 4 | Cameroon | 358 |
| 5 | Congo, Dem. Rep. of | 3 |
| 6 | Congo, Rep. of | 68 |
| 7 | Côte d'Ivoire | 494 |
| 8 | Dominican Republic | 405 |
| 9 | Egypt | 1316 |
| 10 | Gabon | 144 |
| 11 | Guinea | 372 |
| 12 | Honduras | 64 |
| 13 | Indonesia | 558 |
| 14 | Jordan | 213 |
| 15 | Liberia | 64 |
| 16 | Madagascar | 412 |
| 17 | Malawi | 521 |
| 18 | Morocco | 199 |
| 19 | Mozambique | 437 |
| 20 | Nicaragua | 223 |
| 21 | Niger | 816 |
| 22 | Nigeria | 854 |
| 23 | Peru | 615 |
| 24 | Philippines | 350 |
| 25 | Senegal | 570 |
| 26 | Sierra Leone | 12 |
| 27 | Tanzania | 622 |
| 28 | Togo | 114 |
| 29 | Turkey | 225 |
| 30 | Uganda | 358 |
| 31 | Vietnam | 65 |
| 32 | Zambia | 479 |

*Table 1: The frequency of children in utero during SDC year and that also died*

## Figures

[see pdf attached to the main document]
*Figure 1:Ttime lag between a child's birth and its country's SDC*

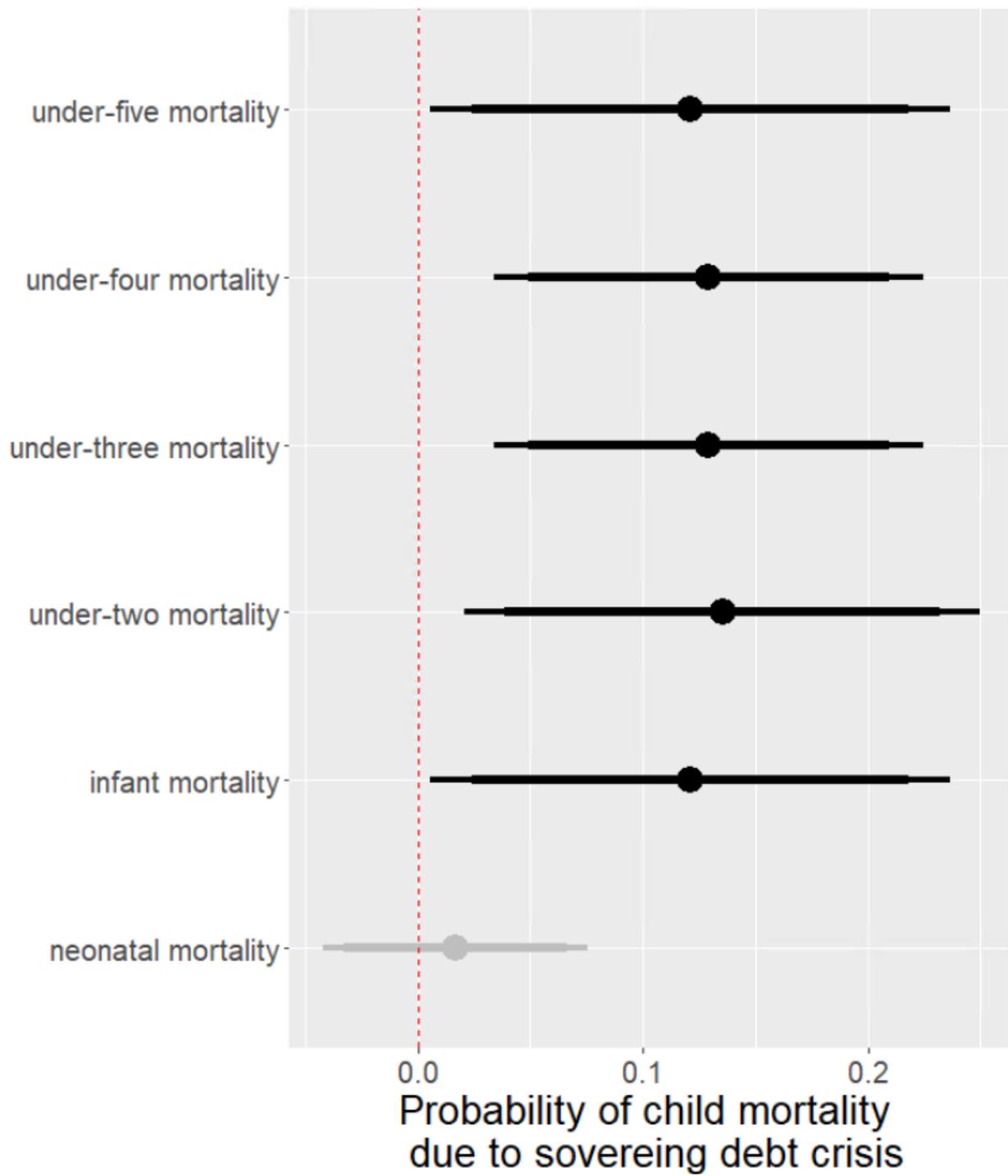

*Figure 2: Average treatment effect*

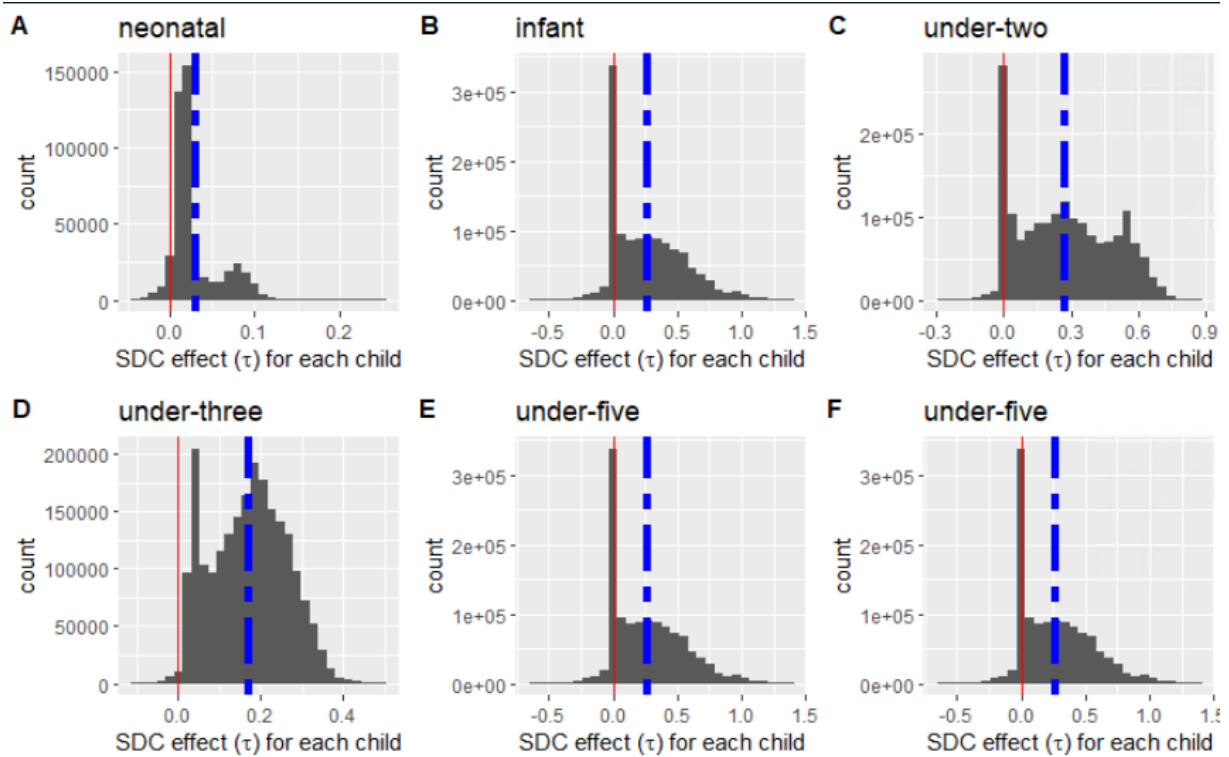

*Figure 3: The distributions of conditional average treatment effects (CATE) by child mortality sample*

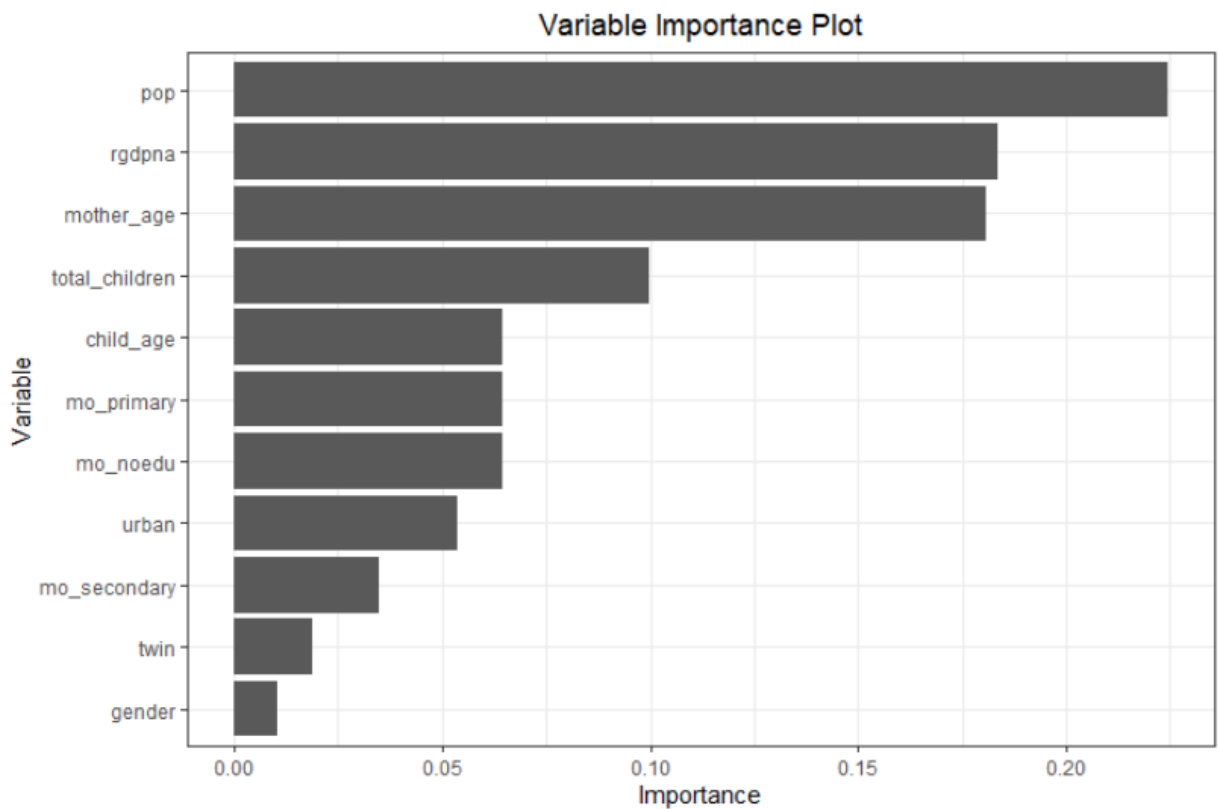

*Figure 4: A ranking of the moderating variables of treatment heterogeneity for neonatal mortality*

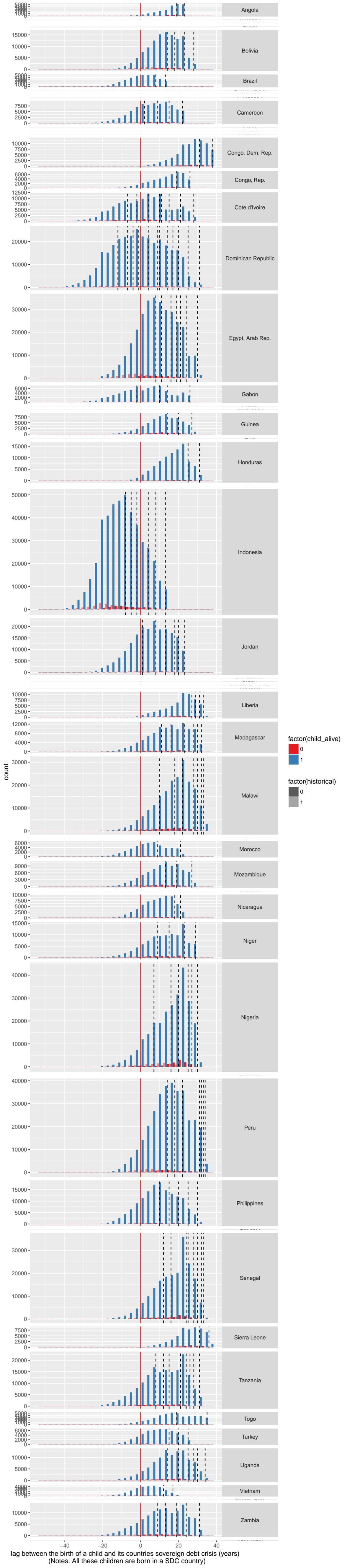

lag between the birth of a child and its countries sovereign debt crisis (years)
(Notes: All these children are born in a SDC country)